\documentclass[aps,prl,twocolumn,superscriptaddress,nofootinbib,
nobalancelastpage,showpacs,nobibnotes]{revtex4}
\usepackage{epsfig}

\begin{document}
\title
{ Nuclear fusion in dense matter
}

\author{R. F. Sawyer}
\affiliation{Department of Physics, University of California at
Santa Barbara, Santa Barbara, California 93106}

\begin{abstract}
The standard theory of nuclear fusion rates in strongly interacting plasmas 
can be (correctly) derived only when the energy release, Q, is large
compared to other energies in the problem. We exhibit a result for rates
that  provides a basis for calculating the finite $Q$ corrections. Crude estimates 
indicate a significant defect in the conventional results for some regions of high
density and strong plasma coupling. We also lay some groundwork for a path integral
calculation of the new effects.


\end{abstract}
\maketitle

The calculations of nuclear fusion rates in strongly coupled equilibrium plasmas, as required for some stellar environments, nearly all
rely on one simple premise, namely that the rate per unit volume $w_s$ of fusion of ions I$_1$ and I$_2$ is given
by, 
\begin{eqnarray}
w_s=n_1 n_2 \langle \sigma v \rangle  K_{1,2}(r=0)/  K_{1,2}^{(0)}(r=0)\,,
\label{assumption}
\end{eqnarray}
where $n_1, n_2$ are the number densities
of the species,
$\sigma$ is the cross-section in vacuum and $K_{1,2}$ is the two body correlator in the presence of the plasma. $K_{1,2}^{(0)}$
is the correlator in the absence of the plasma, and basically cancels out the dependence of the cross-section 
factor on the Coulomb interaction between I$_1$ and I$_2$.
The calculation of $K_{1,2}$ has engendered a big literature, e.g. \cite{gasques} and references
cited therein, in which the correlator is
calculated classically in Monte Carlo simulations at larger distances, leading to an effective two-body potential
defined as $V_{\rm eff}(r)=-\log[K_{1,2}(r)]/ \beta$, where $\beta=(k_B T)^{-1}$. This is followed by a quantum tunneling calculation using this
potential to obtain the overlap at (near) zero separation. This ``basically classical" approach has provided the rates that are actually used in stellar calculations in which the plasma is strongly coupled. It has been tested in
a very small number of quantum path integral calculations \cite{militzer},\cite{ogata}  of the correlator  under some rather specific conditions (very degenerate electrons, one component plasma).
The results appear to be generally supportive of the basically classical approach, at least in some domains \cite{chugunov}.

In the present paper we address the fact that the assumption (\ref{assumption}) that underlies both of
these approaches is not  justified in some domains in which it is currently being used.
A particular fusion reaction
has an energy release $Q$, and it is only when $Q$ is large that (\ref{assumption}) can be established as
a valid approximation, as noted in refs.\cite{bs}\cite{lowell2}. Here we go a step farther both in addressing
the question of how big $Q$ must be in order that (\ref{assumption}) be usable, and in finding
a framework for numerical evaluation in the cases in which it is not. The results are inconsequential
for solar physics, but relevant in denser systems. 

For the case of two ions in and two ions out, ${\rm I_1+I_2\rightarrow I_3+I_4}$ we take the nuclear fusion Hamiltonian as a point coupling, describing the idealized case in which all of the energy dependence of the laboratory cross-section
in the relevant energy range is from Coulomb interactions,
\begin{equation}
H_{\rm nf}=g e^{-iQ t}\int d {\bf r} \psi_3^\dagger ({\bf r},t)\psi_4^\dagger ({\bf r},t) \psi_1 ({\bf r},t) 
\psi_2 ({\bf r},t) +H.C.\,,
\label{hfus}
\end{equation}
where $Q$ is the energy release in the fusion.
Here the $\psi_i$ are nonrelativistic quantum fields that describe creation or annihilation of
the respective ions, in a Heisenberg picture with respect to the complete Hamiltonian. The fields could be Fermi or Bose;
deviation from Boltzmann statistics for the ions is inconsequential.

The remainder of the complete Hamiltonian is taken as $H=H_{1,2}+H_{3,4}+H_{\rm pl}+H_c$
where $H_{1,2}$ and $H_{3,4}$ are the respective kinetic energies plus Coulomb interactions of the  initial
and final systems in the absence of the surrounding plasma, $H_{\rm pl}$ contains all of the kinetic energy
and Coulomb interactions among themselves of the plasma particles, and $H_c$ is the coupling of the fusing particles and the fusion products to the plasma particles.

We refer the reader to ref.\cite{bs} for the derivation of the basic formal rate expression
 based on (\ref{hfus}),
\begin{eqnarray}
w=g^2 \int_{-\infty}^\infty dt \,e^{i Q t}\int d{\bf r}\Bigr\langle \psi_1^\dagger({\bf r},t)\psi_2^\dagger({\bf r},t)
 \psi_3({\bf r},t)\psi_4({\bf r},t) 
\nonumber\\
\times \psi_4^\dagger({\bf 0},0)\psi_3^\dagger({\bf 0},0)\psi_2({\bf 0},0)\psi_1({\bf 0},0)
\Bigr \rangle_\beta~,~~~~~~~~~~~~~~~~~
\label{orig}
\end{eqnarray}
where the notation $\langle...\rangle_\beta$ indicates the thermal average in the medium, 
 $\langle O\rangle_\beta\equiv Z_P^{-1}{\rm Tr} [O\exp(-\beta H)]$ with $Z_P$ the partition function.
We take a one component ionic plasma neutralized by degenerate electrons. 
In this case  the plasma coordinates are the ionic positions ${\bf R}_1....{\bf R}_{N}$. Then
we can transform the result (\ref{orig}) into,
\begin{eqnarray}
w=C\int d{\bf r} \int_{-\infty}^\infty dt ~
e^{i Q t} \int  d{\bf R}_1, ..d{\bf R}_N,..d {\bf R}'_1,..d{\bf R}'_{N}\times&  
\nonumber\\  
 \langle {\bf r}_1={\bf r}_2=0,{\bf R}_1,..{\bf R}_N |e^{-(\beta-i t)  H} | {\bf r}_1
={\bf r}_2={\bf r}, {\bf R}_1',..{\bf R} _N' \rangle_{1,2}&
\nonumber\\
\times \langle {\bf r}_3={\bf r}_4={\bf r},{\bf R}_1',.{\bf R}_N' |e^{-i t H}  |{\bf r}_3={\bf r}_4=0,{\bf R}_1,..{\bf R}_N \rangle_{3,4},&
\nonumber\\
\,
\label{basic}
\end{eqnarray}
where the first two arguments in the kets are the coordinates of the reacting ions in the fusion process and the
coordinates ${\bf R}_i$ stand for all the other ions in the plasma. The redundant subscripts $1,2$ and $3,4$
on the respective brackets $\langle \rangle$ are a reminder of which of the reacting ions
are present in the states within the brackets.
The multiplying constant is $C=Z_P^{-1}n_1 n_2g^2 $.
The individual steps in going from (\ref{orig}) to (\ref{basic}) are simple: first the expression of the Heisenberg fields in terms of Schrodinger
fields, $\psi _S({\bf r})=\exp[-i H t] \psi_H({\bf r},t)\exp[i H t]$, then introduction of the single particle states for the 
respective reacting particles,
$|{\bf r}\rangle=\psi_S^\dagger({\bf r}) |0 \rangle$ and the explicit introduction of the plasma coordinates.

The result (\ref{basic}) is equivalent to the interaction picture result eq. (2.53) of ref. \cite{bs}. The latter
is better suited to perturbation expansions; the former to our present purposes.

To get $w_s$ of the  standard theory as given by (\ref{assumption}) we strike the space and time dependence 
of the first bracket in (\ref{basic}) and take the factor involving the fusion products to be independent
of the plasma coordinates,
\begin{eqnarray}
w_s = C
 \int  d{\bf R}_1, ...d{\bf R}_N ~~~~~~~~~~~~~
\nonumber\\  
 \times \langle {\bf r}_1={\bf r}_2={\bf 0},{\bf R}_1,..{\bf R}_N |e^{-\beta  H}  |{\bf r}_1={\bf r}_2={\bf 0},
{\bf R}_1,..{\bf R}_N\rangle_{1,2}
\nonumber\\
\times \int_{-\infty}^\infty dt \,e^{i Q t}\int d{\bf r} \langle {\bf r}_3={\bf r}_4={\bf r }|e^{-i t H_{3,4} }
 |{\bf r}_3={\bf r}_4=0 \rangle_{3,4}\,.
\nonumber\\
\label{standard}
\end{eqnarray}
In the expression (\ref{standard}) the first two lines give  $n_1 n_2 g^2 K_{1,2}(0)$ as appear 
in (\ref{assumption}).  The final
line in (\ref{standard}) calculates the phase space, $\Omega$, for the fusion products, and the effects of their mutual Coulomb interaction,
in a limit in which their final energy is exactly $Q$ and their total momentum exactly zero. Using $g^2 \Omega=\langle \sigma v\rangle  [K^{(0)}]^{-1}$ we obtain (\ref{assumption}), with the caveat that we have calculated the final phase space
neglecting momentum and energy pass-through from the initial state.
But this is inconsequential in the large $Q$ limit, which is for other reasons the domain of applicability of (\ref{standard}),
as noted in ref. \cite{bs} and explicitly demonstrated in an example below. 

The best calculation to date of the correlator  $K_{1,2}$, implicit in the first factors of  (\ref{standard}), appears to be that by Militzer and Pollack
 \cite{militzer}. Using $ \exp(-\beta H)=[\exp(-\beta H /N_1)]^{N{_1}}$ in (\ref{standard}), when
 $N_1$ is sufficiently
large, they use perturbation theory for the individual factors, each effectively now at high temperature, $N_1T$. This is an expensive
calculation because $N$ and $N_1$ have to be fairly large, and between each of the $N_1$ factors one must integrate over the full manifold of ${{\bf r}_1',{\bf r}'_2,\bf R}_1'...{\bf R}'_N$.

To apply the same technique to the complete formulation, (\ref{basic}), we would face the additional complications
of the time and space integrations, requiring a computation for each point $r,t$ that is the equilivalent 
of the entire calculation of  (\ref{standard}).  Furthermore, as it stands, a calculation of (\ref{basic}) at even one point 
$r,t$
is out of the question because of the oscillating integrands. To make progress, we look at an approximate form of the last bracket $\langle \rangle$ in
(\ref{basic}) for small values of $t$. 
For simplicity we take ions $I_2$ and $I_4$ to be infinitely massive and to be situated at the origin; we can then eliminate their coordinates altogether, and also eliminate all mention of the center of mass position, $r$, in (\ref{basic}). 
In the last bracket $\langle\rangle$ in (\ref{basic}) where the coupling is to the outgoing ions
 \#3 and \#4 we make the further simplification in the factor relating to the
fusion products,
\begin{eqnarray}
 \langle {\bf 0}|e^{-it (H_{\rm pl} +H_{3,4}+H_c)}|{\bf 0}\rangle \approx 
\Lambda 
 \, \langle {\bf 0}|e^{-i t H_{3,4}}|{\bf 0}\rangle\,,
\label{second}
\end{eqnarray}
where
\begin{eqnarray}
 \Lambda= 1-i t H_{\rm pl}-i t (e_3+e_4)\phi({\bf 0})\,.
\label{lambda}
\end{eqnarray}
Here  $\Lambda$ is an operator in the plasma space, with the $R_i$ indices 
now suppressed. The label ${\bf 0}$ in the kets  refers only to the position of I$_1$.
In (\ref{lambda}), $\phi({\bf r })$ is the operator for the electric potential of the plasma,
which enters in the coupling term of the I$_3$, I$_4$ system to the plasma,
$H_C=e_3 \phi ({\bf r_3})+e_4\phi({\bf 0})$. 
Commutators that have been neglected in writing (\ref{second}) in the above form give terms of order $t^2$ and higher, and would give terms of higher order in $Q^{-1}$ than those that we estimate below. 
 
From (\ref{basic}) the rate is now,
\begin{eqnarray}
w=\int_{-\infty}^\infty dt ~
e^{i Q t} F(t) G(t)\,,
\label{rate}
\end{eqnarray}
where 
\begin{eqnarray}
F(t)=C
 {\rm Tr}_{pl}\Bigr[\langle {\bf 0} |e^{-(\beta-it)  H}\Lambda |{\bf 0}\rangle \Bigr ]\,,
\label{F}
\end{eqnarray}
 with the trace performed in the plasma space. $G$ is given by,
\begin{eqnarray}
G(t)=\int {d{\bf q'}\over (2 \pi)^3 } e^{-i {q^2 \over 2M}
(t-i \epsilon )}|\Psi_{\bf q'}({\bf 0})|^2
\nonumber\\
\approx i^{-3/2}(t-i \epsilon )^{-3/2}\Bigr ({ M\over 2\pi}\Bigr)^{3/2}\,,
\label{G}
\end{eqnarray}

 In (\ref{G}),  $\Psi_{\bf q'}({\bf 0})$ is the Coulomb wave function  for asymptotic momentum $q'$, evaluated
at the origin. In the second line we have discarded the Coulomb interaction between $I_3$ and  $I_4$.
The function $G(t)$ is analytic in the upper half $t$ plane except for the branch point at $i \epsilon$; note that
this singularity comes from the high $q^2$ behavior of the integrand in (\ref{G}) and should have the 
same form when we restore the Coulomb force between the final particles. 
A similar analysis of $F(t)$, leaving the plasma out entirely, would give  a structure with a singular
factor $(t+i \beta)^{-3/2}$. We assume that this analytic structure persists in the presence of the plasma.

Then defining the cut in the $t^{-1/2}$ function in (\ref{F}) to run from $i\epsilon$ to $i \infty$,  and that in 
$(t+i \beta)^{-3/2}$ in $F$ to run from  $-i\beta$ to $-i \infty$, we deform  the $t$ integration
contour to run from $i \,\infty-\epsilon$ to $0$ to $i\, \infty+\epsilon$ and  replace it with a simple integral
of the discontinuity and a real integration variable, $\tau$,
ending with,

\begin{eqnarray}
&w=i^{-3/2}\int_{- \infty}^\infty dt\, F(t)e^{i Q t} (2\pi M)^{3/2}(t-i \epsilon )^{-3/2}
\nonumber\\
&=2 i^{-3/2}\int_{- \infty}^\infty dt\, {d \over dt}[e^{i Q t}F(t)] (2\pi M)^{3/2}(t-i \epsilon )^{-1/2}
\nonumber\\
&= -2^{1/2}\Bigr({ M\over  \pi }\Bigr)^{3/2} \int_0^\infty d\tau ~\tau^{-1/2}\, {d \over d \tau}\Bigr [
e^{-Q \tau} F( i \tau)\Bigr]\,,
\label{crazy}
\end{eqnarray}
the integration by parts before transforming the contour being required in order to avoid a 
non-integrable singularity at $t=0$ in the final form.

The rate corrections of order $Q^{-1}$ are now calculated from the linear term in the expansion of
of $F(t)$ in powers of $t$. From (\ref{F}) and (\ref{lambda}) we obtain just,
\begin{eqnarray}
F(t)= 
 C{\rm Tr}_{pl}\Bigr[\langle {\bf 0} |e^{-\beta  H} [1+i t H_{1,2}] | 
{\bf 0}  \rangle_{1,2}\Bigr ]\,,
\label{F1}
\end{eqnarray}
the terms with $t\phi(0)$ having cancelled through conservation of charge, and the terms with $t H_{\rm pl}$ having cancelled as well.
The term with unity in the final bracket gives the rate $w_s$ of 
(\ref{standard}). We estimate the contribution $\delta F$ of the linear term in $t$ the ``basically classical" approximation, in which plasma coordinates are eliminated in favor of an effective Hamiltonian $H_{\rm eff}={\rm KE}+V_{\rm eff}$,
\begin{eqnarray}
\delta F(i \tau)=
n_1 n_2 g^2 \zeta^{-1}\tau \langle {\bf 0}|e^{-\beta H_{\rm eff} }H_{1,2}|{\bf 0}\rangle\,,
\label{crazy3}
\end{eqnarray}
where $\zeta=\int d {\bf r} \langle {\bf r}| e^{-\beta H_{\rm eff}}| {\bf r}\rangle$.
We evaluate (\ref{crazy3}) taking  a minimal modification of the potential due to the plasma, following 
ref. \cite{aj}, 
\begin{eqnarray}
\delta H\equiv H_{\rm eff}-H_{1,2}\approx -\Gamma \beta^{-1}[1-(r/2a)^2]\,,
\label{veff}
\end{eqnarray}
where $a$ is the average interparticle spacing and $\Gamma$ the classical plasma coupling strength.
 We evaluate using
\begin{eqnarray}
\langle {\bf 0}|H_{1,2}e^{-\beta H_{\rm eff}}|{\bf 0}\rangle=-\langle {\bf 0}|
[{\partial\over \partial \beta}
+\delta H]e^{-\beta H_{\rm eff}}|{\bf 0}\rangle\,.
\label{expect}
\end{eqnarray}
and
\begin{eqnarray}
e^{-\beta H_{\rm eff} }\approx e^{-S_0}e^\xi\,,
\end{eqnarray}
where  $S_0=[27 \pi^2 \beta M (Ze)^4 /4 \hbar^2]^{1/3}$  and
$\xi= [\Gamma-(45 \Gamma^3 )/ (32 S_0^2)]$, as in eq.28 of ref.\cite{aj}. Here $\Gamma=Ze^2\beta/a$ where 
$a=(3/4 \pi n_I)^{1/3}$. In the second term in (\ref{expect}) we
take $\delta H $ at the 
time averaged (imaginary) time $\bar r$, eq.13 in ref.\cite{aj}.
We obtain,
\begin{eqnarray}{\delta w \over w_s}=
 \,-{\partial S_0 \over \partial\beta}Q^{-1}-{129 \over 64} {\Gamma^3\over S_0^2 \beta Q}\,.
\label{crazy4}
\end{eqnarray}

 The first term on the RHS of (\ref{crazy4})
simply adjusts the phase space for the outgoing particles by adding an energy on
the order of the Gamow energy to the final state,  note the discussion of energetics
above, after (\ref{standard}). 
The remaining term comes specifically from the $r^2$ term in the effective potential. (The
contributions from the $r$ independent modification canceled, using $(\partial /\partial \beta )\Gamma =
\Gamma/\beta$.) This $r^2$ term gives a fractional correction  
$\approx .4 \,{\rm MeV}/Q$ for the case of
the extreme conditions of 
 $^{12}$C +$^{12}$C at a temperature of $10^8$K  and a density of $10^{10}{\rm g c}^{-3}$.

A correction of the above magnitude does not necessarily create a problem  for, say, the conventional calculation of $^{12}{\rm C }+^{12}{\rm C}\rightarrow ^{23}{\rm Na +p }$ where the $Q$ value is about 2 MeV, and where the applications care about factors of ten and not about 20\% corrections. But the fractional correction is much larger than
$T/Q$. 
The next
logical step would be to expand of the right hand $\langle \rangle$ in 
(\ref{basic}) in an infinite series of operators in the plasma space, of which (\ref{F1}) displays the first two terms.
Our conjecture is that the contour distortion can then be applied to give individual terms each of which
could be calculated with the path integral method.
 
We can get complementary information from perturbative calculations in a system in which the expansion in powers 
of $Q^{-1}$ is not applicable.
We consider  a reaction I$_1$+I$_2\rightarrow$ I$_3$, where
 I$_3$ is a narrow resonance, and where the 
decay of the resonance is nearly all into channels other than the entrance channel. 
The formalism above is applicable simply by eliminating every 
reference to ion \#4 in every equation. The local interaction, (\ref{hfus}), removing the $\psi_4^\dagger$
factor,
perfectly describes the limit in which the Breit-Wigner formula, with Coulomb removed, becomes
a constant times a delta function in energy.
As context we mention the possible application to  $^{12}$C +$^{12}$C  fusion at temperatures in the range
of a few times $10^8$ K, where the magnitude of $Q$, which is \underline{negative} in this case, is chosen to be 
close to the Gamow peak for the fusion reaction, resulting in enhancements to the rate \cite{cussons}-
\cite{cooper}.

We consider only the order in which there are two interactions of the distinguished ions $I_1, I_2, I_3$
with the plasma; the superficial order of $w$ in the coupling parameter is $e^4$ but
the long range part of the Coulomb force reduces the order of the leading term
to $e^3$. For these terms the dimensionless
strength parameter is $\lambda_1=e^2 Z^2 \kappa_D$ with $\kappa_D^2=4 \pi n_I Z^2 \beta$.
To calculate we introduce an interaction representation in which the ``interaction'' Hamiltonian is
$H_c$, the coupling of $I_1,I_2 ,I_3$ to the plasma particles, and the unperturbed Hamiltonian, $H_0=H-H_c$.
The calculations are fairly standard, with the replacement, e.g. within the first $\langle\rangle$ in (\ref{basic}),
\begin{eqnarray}
e^{-H (\beta-it)}=e^{-H_0(\beta-i t)} \exp\Bigr [i \int_{-i \beta}^t dt' H_c^I(t')\Bigr ]_+\,,
\label{int}
\end{eqnarray}
where the final subscript indicates time ordering along the path $-i \beta $ to $0$ to $t$. The perturbation terms
will come from the first and second terms in the expansions of the final exponentials of (\ref{int}) and its
(anti-time-ordered) counterpart for the second $\langle\rangle$ in (\ref{basic}).
To distinguish the source of the terms that follow we introduce independent charges of the fusing particles and the fusion
products,
$e_1,e_2,e_3$, taking $e_1=e_2=eZ=e_3/2$ at the end of the calculation.
When we calculate just the terms of order $e^3$ in (\ref{basic}) we obtain,
\begin{eqnarray}
&w=g^2 n_1 n_2 \zeta^{-1}\int_{-\infty}^{\infty}dt 
[1-\beta^{-1}\kappa_D J(t)] e^{i Q t} \int d{\bf r}
\nonumber\\
 &\times
\langle {\bf 0},{\bf 0}|e^{-H_{1,2}(\beta-i t)}
|{\bf r},{\bf r}\rangle_{1,2}
 \langle {\bf r}|e^{-i H_{3}t} |{\bf 0},{\bf 0}\rangle_3\,,
  \label{xsal}
\end{eqnarray}
where
\begin{eqnarray}
&J(t)= (e_1+e_2)^2({t^2\over 2}-{\beta^2\over 2}+i\beta t)+
e_3^2 \,\,{t^2\over 2}
\nonumber\\
 &-(e_1+e_2) e_3  (t^2+i \beta t)\,,
\label{timeint}
\end{eqnarray}
 When charge conservation is imposed,  the individually time dependent terms in (\ref{timeint}) add up  to
 $-(e_1+e_2)^2 \beta^2/2$, and we obtain  the perturbative change in rate,
\begin{eqnarray}
\delta w= (1/2) (e_1+e_2)^2\kappa_D \beta w^{(0)}\,,
\label{ans}
\end{eqnarray}
 where however the $e_1^2$ and $e_2^2$ terms are to be dropped because they are exactly compensated by the changes in chemical potential
required to keep the number densities of the respective species constant, as shown explicitly in ref \cite{bs}. In the end 
we obtain just the
Salpeter \cite{salpeter} correction (for $Z_1=Z_2=Z$), $\delta w=\lambda_1 w^{(0)}$.  Using instead just the term in (\ref{timeint}) in which the plasma couples only to the incoming ions would produce a spurious addition of relative order $(\beta Q)^{-1}$ to our earlier large $Q$ expansion, and more damaging spurious effects in a resonance
case (where we cannot use the $Q^{-1}$ expansion).

Deriving the correct Salpeter result
from a wrong equation, (\ref{standard}), can lead one to believe that there exist further corrections of order $\lambda_1$
coming from the coupling of the fusion products to the plasma. Other works, e.g. refs.\cite{itoh} \cite{lamb},
have combined a standard screening enhancement, of order $e^{\Gamma}$, with an resonance energy shift,
of order $\Gamma/\beta$. In our weak coupling framework, including the latter would be incorrect, since the
fractional correction to order $\lambda_1$ is given completely by (\ref{ans}). Thus it appears to be incorrect
in the strongly coupled case as well, though the issue deserves further scrutiny.

We have calculated the residual, non-infrared divergent, terms of order $e^4$ in the perturbation
expansion for the resonance production case coming from two interactions of the plasma with the final 
resonance.
We find a fractional correction of a few times $\lambda_2 L[Q/E_{\rm Gamow}]$ where 
$\lambda_2 =e^4 Z^3 \hbar \beta^{5/2}n_I M^{-1/2}$ and the function
$L$ is of order unity when $Q$ is of order $E_{\rm Gamow}$. Thus the correction is superficially of
order $\hbar$ while the order $\lambda_1$ correction of (\ref{ans})is classical. However, there is also 
implicit $\hbar$ dependence in $E_{\rm Gamow}$.
For example in  $^{12}$C+$^{12}$C at a temperature of $ 10^8 K$ , this correction would be  100\% for densities greater than about $10^8 {\rm g c}^{-3}$. The classical coupling $\Gamma$ (or $\lambda_1^{2/3}$)
has become strong for even lower density, so this perturbative estimate of a quantum effect must be regarded
with suspicion, and is of no value to phenomenology. But it is worrisome that the usual picture
does not in any way include the physics of these terms,  especially when applied at one hundred times the density
at which they appear to become important. 

The most important new results in this paper are the presentation of the basic governing 
 equation (\ref{basic}) and the demonstration that the real time dependence therein can lead to big changes
when $Q$ is not sufficiently large. It is noteworthy that, even when we evaluated the correction using
the classical potential, the time dependence in the framework combines with the quantum effects in the short 
distance tunneling region  to produce significant changes. The last, perturbative, parts of the paper  
serve as a further caveat with respect to using the existing lore when $Q$ is not large.

The author owes thanks to Randall Cooper for getting him interested in the plasma corrections in the
resonance case, and to Lowell Brown for teaching him much of the underlying physics.
 This work was supported 
in part by NSF grant PHY-0455918. 

\end{document}